\documentclass[aps,prl,twocolumn,showpacs]{revtex4}
\usepackage{amsfonts}
\usepackage{amsmath}
\usepackage{graphicx}
\usepackage{dcolumn}

\begin{document}

\title{\textit{Supersolitons}: Solitonic excitations in atomic soliton chains%
}
\author{David Novoa$^{1}$, Boris A. Malomed $^{2}$, Humberto Michinel$^{1}$, 
and V\'{\i}ctor M. P\'{e}rez-Garc\'{\i}a$^{3}$.}
\affiliation{$^1$\'Area de \'Optica, Facultade de Ciencias, Universidade de Vigo, As
Lagoas s/n, Ourense, E-32004 Spain. \\
$^{2}$ Department of Physical Electronics, School of Electrical Engineering,
Faculty of Engineering, Tel Aviv University, Tel Aviv 69978, Israel \\
$^3$ Departamento de Matem\'aticas, E. T. S. I. Industriales, and Instituto
de Matem\'atica Aplicada a la Ciencia y la Ingenier\'{\i}a, Av. Camilo
Jos\'e Cela, 3, Universidad de Castilla-La Mancha, 13071 Ciudad Real, Spain.}

\begin{abstract}
We show that, by appropriately tuning physically relevant
interactions in two-component nonlinear Schr\"{o}dinger equations,
it is possible to achieve a regime with particle-like solitonic collisions. This allows us to construct an analogue of
the Newton's cradle and also to create localized collective excitations in
solitary-wave chains which are quasi-integrable solitons, i.e. supersolitons.
 We give a physical explanation of the
phenomenon, support it with a perturbative analysis, and confirm our
predictions by direct simulations.
\end{abstract}

\maketitle

\emph{Introduction and model.-} One of the most successful concepts
of nonlinear science with applications to a great variety of
physical contexts is that of solitons, i.e., self-localized
nonlinear waves, sustained by the balance between dispersion and
nonlinearity. Many types of solitons have been
studied, from the classical examples found in integrable models,
such as the Korteweg - de Vries, sine-Gordon, Toda-lattice (TL),
nonlinear Schr\"{o}dinger, and other celebrated equations, and
extending into the realm of realistic non-integrable nonlinear-wave
models.

Solitons are usually expected to be robust against collisions, which is a
trademark feature of integrable equations. A lot of activity has been
directed at the study of soliton collisions and interactions in
non-integrable systems. Recent advances include the analysis of chaotic
scattering  \cite{Nbounce}, the formation of soliton bound states
and soliton clusters \cite{soliclus}, and the studies of
soliton collisions in vector systems \cite{Nbounce,Gerdjikov,v1}, to name
just a few.

While it is customary to speak of solitons as elastically colliding
quasi-particles, most solitons, specifically in integrable systems, pass
through each other, thus clearly featuring their wave nature. On the other
hand, elastic collisions between classical particles lead to momentum
exchange between them, and rebound, due to the non-penetrability of
classical particles.

In this paper we discuss a particular soliton collision scenario of physical
relevance, where truly elastic particle-like soliton collisions can be
achieved. We will show how this can be used to build a vector-soliton
version of the \emph{Newton's cradle}, and to build \emph{supersolitons},
i.e. collective soliton-like excitations in arrays of solitary waves,
leading to a remarkable conjunction of emergent phenomena: the former one
representing the formation of robust soliton trains, and the latter effect
implying the emergence of an effectively quasi-discrete soliton at a higher
level of organization.

The basic model which allows to implement the above-mentioned effects is
based on the two-component (vectorial) nonlinear Schr\"{o}dinger equation
(NLSE), that arises in sundry contexts \cite{Kiv}. Its normalized form is
\begin{equation}
i\frac{\partial u_{j}}{\partial t}=-\frac{1}{2}\frac{\partial u_{j}}{%
\partial x^{2}}+V(x)u_{j}+\sum_{k=1,2}g_{jk}|u_{k}|^{2}u_{j}.  \label{GP}
\end{equation}%
An important physical realization of this model is a multicomponent
Bose-Einstein condensate (BEC), where $u_{j}$ are wave functions of two
atomic states under the action of a strong transverse trap with frequency $%
\nu _{\perp }$ \cite{PG98}. The variables $x$ and $t$ are measured,
respectively, in units of $a_{0}=\sqrt{\hbar /m\nu _{\perp }}$ and $1/\nu
_{\perp }$, and $g_{ij}\equiv 2a_{ij}/a_{0}$, with $a_{ij}$ the respective $%
s-$wave scattering lengths. The normalization integral for $u_{j}$ gives the
number of atoms in the respective species, $\int_{-\infty }^{+\infty
}|u_{j}|^{2}d^{3}x=\mathcal{N}_{j}$. Solitons in BECs 
%with attractive interactions between atoms 
have been created experimentally \cite{solibec}
and their interactions studied theoretically in many papers (see e.g. \cite%
{coli1,Gerdjikov2,Adams}).

\emph{Particle-like elastic collisions and the solitonic Newton's cradle.-}
We will consider  Eqs. \eqref{GP} with intra-component
attraction ($g_{11},g_{22}<0$) and inter-component repulsion ($%
g_{12},g_{21}>0$). To fix ideas we will choose $g_{11}=g_{22}=-g_{12}=-g_{21}(=1$ without loss of generality)
 for which case the coupled NLSE are not integrable \cite%
{Schulman}. In this situation the solitons, that may be formed in both
components independently, interact incoherently with a repulsive force. The basic physical feature underlying our analysis is that the dynamics of those
solitons may be similar to that of elastic beads. We will explore the cases of
 harmonic longitudinal confinement $V(x)=\Omega ^{2}x^{2}/2$, and
ring-shaped configurations \cite{RT1}. 

\begin{figure}[tbph]
{\centering
\resizebox*{1\columnwidth}{!}{\includegraphics{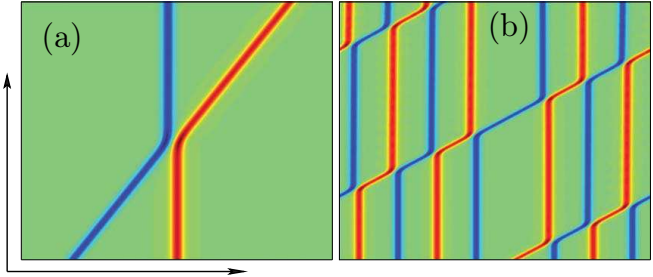}} }
\caption{[Color online] Pseudocolor plots of $|u_{1}(x,t)|^{2}$ (blue) and $%
|u_{2}(x,t)|^{2}$ (red) show the time evolution of initially well separated
soliton trains built as per Eq. (\protect\ref{initia}). The horizontal
(vertical) axis of each plot is spatial adimensional variable $x$
(adimensional time $t$). (a) Collision of two solitons ($N=1$) for $\protect%
\xi _{1}=-20,\protect\zeta _{1}=0,v_{1}=0.4,w_{1}=0$, $x\in \lbrack -30,30]$
and $t\in \lbrack 0,100]$. (b) Multisoliton collisions in a ring for $N=4$, $%
\protect\zeta _{j}=-35+20j,\protect\xi _{j}=-25+20j$, $x\in \lbrack -40,40]$
and $t\in \lbrack 0,250]$. Initial velocities are all zero except for $%
w_{3}=0.5$.}
\label{fig1}
\end{figure}

Let us consider soliton trains built as follows:
\begin{subequations}\label{initia}
\begin{eqnarray}
u_{1}(x,0)&=&\sum_{n=1,...,N}(-1)^{n}\text{sech}\left( x-\xi _{n}\right) \exp
\left( ixv_{n}\right) , \\
u_{2}(x,0)&=&\sum_{n=1,...,N}(-1)^{n}\text{sech}\left( x-\zeta _{n}\right) \exp
\left( ixw_{n}\right) ,%
\end{eqnarray}%
\end{subequations}
with alternation of the soliton species in the train, i.e.,%
\begin{equation}
...\xi _{n-1}<\zeta _{n-1}<\xi _{n}<\zeta _{n}<\xi _{n+1}<\zeta _{n+1}<...,
\label{ordenar}
\end{equation}%
$v_{n}$ and $w_{n}$ being initial velocities of the solitons. In Fig. \ref%
{fig1}, where the trap is absent, panel (a) displays a single collision
event ($N=1$). It is noteworthy that, because of the repulsive
inter-component interaction, the incident soliton (in field $u_{1}$)
transfers all of its momentum to the initially static soliton (in $u_{2}$),
in full compliance with the behavior of elastic particles and contrary to
the typical behavior of nontopological solitons in integrable systems. The
dynamics of a train of eight alternating solitons in a ring configuration
demonstrates the periodic transfer of the momentum through the train, see
Fig \ref{fig1}(b).

\begin{figure}[tbph]
{\centering
\resizebox*{1\columnwidth}{!}{\includegraphics{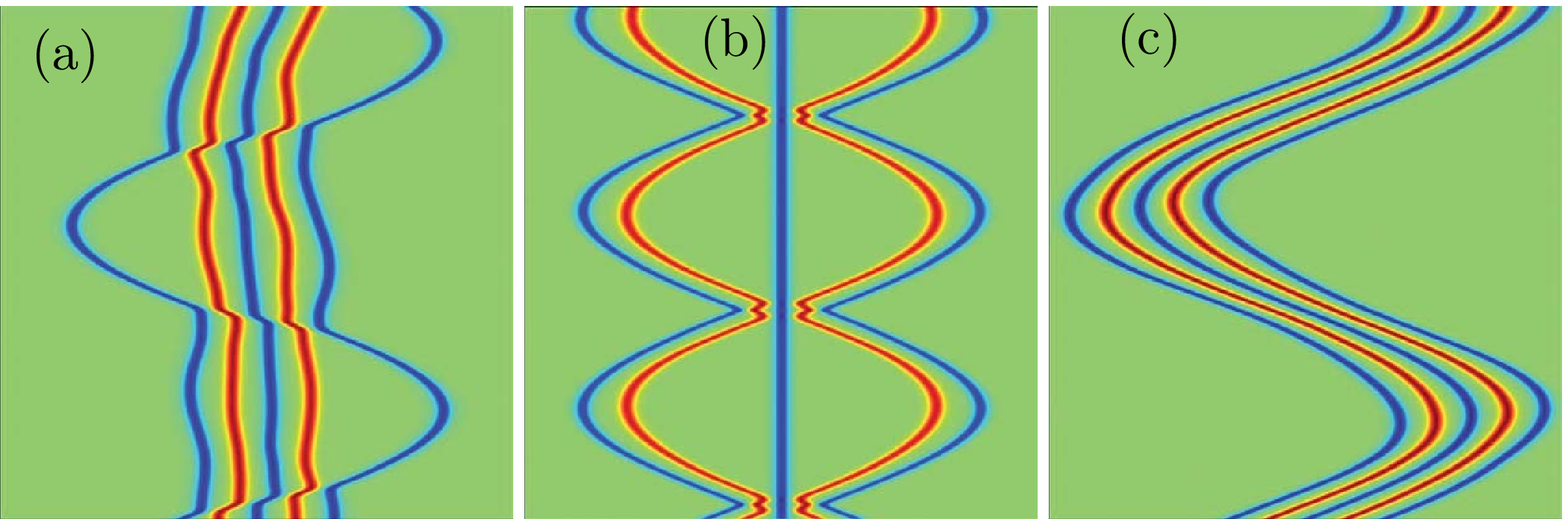}} }
\caption{(Color online) Same as Fig. {\protect\ref{fig1}} for a \emph{%
Newton's cradle} built up of five solitons with an external
potential with $\Omega ^{2}=6\cdot 10^{-5}$. In all subplots $t_{\mathrm{max}%
}=1000$, $x\in \lbrack -40,40]$, and the initial separation between solitons
is $\protect\xi _{n}-\protect\zeta _{n}=6$. Different oscillation modes are
excited by applying an initial velocity, of size $0.2$, to: (a) the top-left
soliton, (b) two ultimate solitons on both sides, with opposite velocities,
and (c) all solitons. }
\label{fig2}
\end{figure}

By adding the parabolic trapping potential we urge the solitons to oscillate
around the equilibrium position. In the quantum interpretation of our model
this setup would provide a way to construct a \emph{quantum Newton's cradle}
with atomic solitons, as illustrated in Fig. \ref{fig2}. Unlike other
settings explored in BEC \cite{cradd}, the cradle configuration does not
require a lattice potential to create effective particles, which are here
created by purely nonlinear interactions.

\emph{The Toda-lattice limit: supersolitons -} We now consider two
alternating chains of solitons set along a ring. Within the
effective-particle approach, we assume that all solitons, which have
identical amplitudes in each component, $\eta $ (for $u_{1}$) and $\theta $
(for $u_{2}$), behave like rigid particles and thus do not suffer
conspicuous deformation, i.e., each soliton may be approximated by
\begin{subequations}
\begin{eqnarray}
u_{1}^{(n)}(x,t) &=&\eta\, \mathrm{sech}\left[ \eta \left( x-\xi _{n}\right) %
\right] 
e^{i\dot{\xi}_{n}x+\frac{1}{2}\int \left( \eta ^{2}-\dot{\xi%
}_{n}^{2}\right) dt}, \\
u_{2}^{(n)}(x,t) &=&\theta\, \mathrm{sech}\left[ \theta \left( x-\zeta
_{n}\right) \right]
e^{i\dot{\zeta}_{n}x+\frac{1}{2}\int \left(
\theta ^{2}-\dot{\zeta}_{n}^{2}\right) dt} ,
\end{eqnarray}%
with the initial positions arranged as per Eq. (\ref{ordenar}).

A straightforward analysis based on the perturbation theory for solitons
\cite{PS} yields the following system of equations of motion for the soliton
coordinates:
\end{subequations}
\begin{subequations}
\label{TL}
\begin{eqnarray}
\ddot{\xi}_{n} &=&-\Omega ^{2}\xi _{n}+8\theta \left[ \eta ^{2}\left(
e^{-2\eta \left( \xi _{n}-\zeta _{n-1}\right) }-e^{-2\eta \left( \zeta
_{n}-\xi _{n}\right) }\right) \right.   \notag \\
&&\left. +\theta ^{2}\left( e^{-2\theta \left( \xi _{n}-\zeta _{n-1}\right)
}-e^{-2\theta \left( \zeta _{n}-\xi _{n}\right) }\right) \right] , \\
\ddot{\zeta}_{n} &=&-\Omega ^{2}\zeta _{n}+8\eta \left[ \eta ^{2}\left(
e^{-2\eta \left( \zeta _{n}-\xi _{n}\right) }-e^{2\eta \left( \xi
_{n+1}-\zeta _{n}\right) }\right) \right.   \notag \\
&&\left. +\theta ^{2}\left( e^{-2\theta \left( \zeta _{n}-\xi _{n}\right)
}-e^{-2\theta \left( \xi _{n+1}-\zeta _{n}\right) }\right) \right] .
\end{eqnarray}%
These equations are derived under the assumption that adjacent solitons are
well separated, i.e., $\left( \eta ,\theta \right) \left( \xi _{n}-\zeta
_{n-1}\right) ,\left( \eta ,\theta \right) \left( \zeta _{n}-\xi _{n}\right)
\gg 1$, although a strong inequality is not really necessary here. Similar
ideas have been used to derive equations for the interaction of other
elementary nonlinear structures, which gives rise to different equations at
a higher level of organization \cite{spr,Gerdjikov,Adams}. Equations (%
\ref{TL}) with $\Omega =0$ reduce to the so-called diatomic TL,
which is \emph{not} integrable, although some solutions are known.
%\cite%{review}.

With $\eta =\theta $ and $\Omega =0$ in Eqs. (\ref{TL}) and defining $%
q_{2n}(t)=2\eta \xi _{n}(t),q_{2n+1}(t)=2\eta \zeta _{n}(t)$ and $\alpha
=32\eta ^{4}$, we arrive at the integrable TL model \cite%
{Toda},
\end{subequations}
\begin{equation}
\ddot{q}_{n}=\alpha \left[ e^{-\left( q_{n}-q_{n-1}\right) }-e^{-\left(
q_{n+1}-q_{n}\right) }\right] .  \label{ITL}
\end{equation}%
This model describes a dynamical lattice with the exponential potential of
the nearest-neighbor interaction. However, potentials of the interaction
between adjacent atoms in real condensed-matter systems are never
exponential, being closer to those of nonlinear anharmonic oscillators.
%(which
%gives rise to more realistic models, such as the Fermi-Pasta-Ulam
%lattice \cite{FPU}).
This is why the only experimental realization of the integrable TL
was realized in electric transmission lines \cite{Tom}, that may be readily
designed in exact correspondence to Eqs. (\ref{ITL}). Our
analysis suggests a possibility to create Toda solitons, of both
mono- and diatomic types, as excitations in interwoven arrays of
multicomponent NLSE\ solitary waves.

We name these excitations \emph{supersolitons} since they occur on
top of an array of ``elementary" solitons, and are expected to be as
robust as solitons in integrable models. The same name was
previously applied to solitons in supersymmetric models
\cite{supersym}, and, in a completely different context, to
localized topological collective excitations in chains of fluxons
trapped in periodically inhomogeneous
Josephson junctions  and in layered superconducting structures \cite%
{spr}. The appearance of TL supersolitons represents a remarkable
phenomenon at a higher-organization level, using, as building blocks,
solitary waves of the multicomponent NLSE, i.e. a strongly \textit{%
nonintegrable} model.

In the monoatomic lattice, Eq. (\ref{ITL}) has an obvious equilibrium
solution with $q_{n}=L/\left( 2N\right) $, where $N$ is the number of
solitons in each subchain, and $L$ the total length of the system. For small
perturbations with frequency $\omega $ and wavenumber $k$ around this
configuration, the dispersion
relation is $\omega ^{2}=128\eta ^{4}e^{-\eta L/N}\sin ^{2}(k/2)$. With
respect to the quantization imposed by the boundary conditions for the
ring-shaped soliton chain, $k=\pi m/N$, $m=0,\pm 1,\pm 2,...$ , this yields%
\begin{equation}
|\omega _{m}|=8\sqrt{2}\eta ^{2}\exp \left( -\eta L/2N\right) \left\vert
\sin (\pi m/(2N))\right\vert .  \label{||}
\end{equation}

If a wave in the lattice is excited by kicking one soliton and lending it
velocity $v$, the wave will hit solitons with period $T=L/(2Nv)$, which
corresponds to an effective excitation frequency $\omega _{\mathrm{exc}}\equiv
2\pi /T=4\pi Nv/L$. Thus, resonant excitations may be expected under the
condition $P\omega _{\mathrm{exc}}=Q|\omega _{m}|$, or, in other words, at
values of the kick velocity belonging to the following \textit{resonant
spectrum},
\begin{equation}
\left\vert v_{m}^{(P,Q)}\right\vert =\frac{Q}{P}\cdot \frac{2\sqrt{2}L\eta
^{2}}{\pi N}e^{-\eta L/2N}\left\vert \sin \left( \pi m/(2N)\right)
\right\vert ,
\end{equation}%
where integers $Q$ and $P$ stand for the order of the resonance and
subresonance ($P=Q=1$ correspond to the fundamental resonance). Another
interpretation of this resonance condition (cf. Ref. \cite{Matteo}) is that
the kick velocity coincides with the phase velocity of linear waves.

\begin{figure}[tbph]
{\centering \resizebox*{1\columnwidth}{!}{\includegraphics{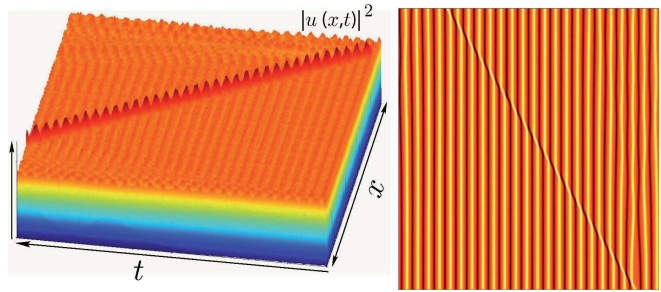}} }
\caption{(Color online) Excitation of a supersoliton on a soliton chain given by Eq. (\ref{initia}-\ref{ordenar}) with $N=24$ localized solitons in each component, by kicking the $j=22$soliton with velocity $w_{22}=-0.5$. Individual solitons have initial amplitudes $\eta = \theta=1$ and the intersoliton distance is five units leading to a total ring length of 240.  Left: Amplitude plot of $|u(x,t)|^2 = |u_1(x,t)|^2 + |u_2(x,t)|^2$. Right: Pseudocolor plot showing $|u_1(x,t)|^2$ (yellow) and $|u_2(x,t)|^2$ (red). The range of adimensional times spanned in both plots is $t\in[0,250]$.}
\label{fig3}
\end{figure}

\emph{Numerical studies of supersolitons -} To verify our predictions based on Eq. (\ref{ITL}), we
have performed numerical simulations of Eq. (\ref{GP}). First, in Fig. \ref{fig3} we have generated a single supersoliton
by kicking one of the most external solitons in one of the components. Since this excitation 
does not correspond exactly to a supersoliton
we also obtain a small ammount of radiation which is seen as small remnant oscillations of the 
individual solitary waves. Apart from this efect due to the excitation 
procedure, the propagation of the supersoliton is perfect as seen both in the amplitude [Fig \ref{fig3}(a)] 
and pseudocolor [Fig. \ref{fig3}(b)] plots. Another effect seen in Fig. \ref{fig3}(a) and not considered 
in our model is the small compression of the individual solitary waves when they are hit by the
 supersoliton (the model assumes equal amplitude individual solitary waves). However this small 
 effect does not affect our conclusions and can be minimized by considering smaller energy collisions (i.e. incident speeds). 

\begin{figure}[tbph]
{\centering \resizebox*{1\columnwidth}{!}{\includegraphics{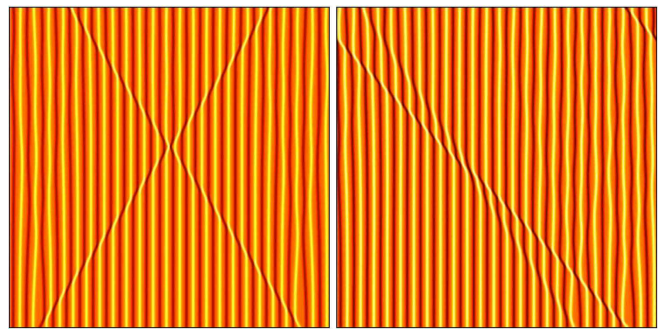}} }
\caption{(Color online) Examples of supersoliton collisions. The initial configuration is as in Fig. \protect\ref{fig3} but 
now two supersolitons are excited. Left: Head-on collision induced with excitation parameters $v_3=0.5$ and $w_{22}=-0.5$. 
The range of adimensional times spanned is $t\in[0,250]$.
Right: Soliton overtaking excited with initial velocities $v_{23}=-0.6$, $w_{19}=-0.3$. 
The range of adimensional times spanned is $t\in[0,350]$.}
\label{fig4}
\end{figure}

Fig. \ref{fig4} shows the collisional behavior for head on collisions of equal speed solitons [Fig. \ref{fig4}(a)] and the overtaking of a slow supersoliton by a faster one [Fig. \ref{fig4}(b)]. In both cases the supersolitonic excitations behave as true solitons, what it is justified by the integrability of our simple model given by Eqs. (\ref{ITL}). We want to emphasize again that these behaviors, typical of integrable systems, arise on top of a strongly nonintegrable model.

\begin{figure}[tbph]
{\centering
\resizebox*{1\columnwidth}{!}{\includegraphics{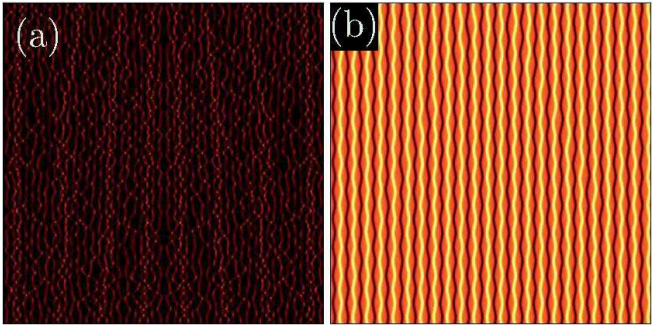}} }
\caption{(Color online) Time evolution for $0<t<500$ of matter-wave soliton
trains with a small collective excitation in the case of one (a) and two (b)
species, which are described, respectively, by the scalar and coupled NLSEs
(black and white colors). Adjacent solitons are given opposed input
velocities, $\protect\chi =\pm 0.1$ in (a), and $v_{j}=0.1$, $w_{j}=-0.1$ in
(b). Other parameters are as in Fig. \protect\ref{fig3}.}
\label{fig5}
\end{figure}

\emph{Can scalar models support supersolitons?-} Soliton collisions
in the framework of scalar NLSEs have been studied in various
contexts , and equations similar to Eqs. (\ref{TL}) have been
derived using different approaches \cite{Gerdjikov,Adams}, leading
to the so-called complex TL. Despite the formal similarities, the
ensuing dynamics is not robust, and solitonic solutions turn out to
be \emph{unstable} because of the phase
dependence of the interactions. An example is displayed in Fig. \ref{fig5}%
(a), which shows that the phase shifts induced by the initial kick velocity
lead to an unstable dynamics of the single-component chain, whereas in its
alternating two-component counterpart the system does not display any
instability, as shown in Fig. \ref{fig5}(b); in particular, the
configuration shown in Fig. \ref{fig5}(b) periodically recovers its shape.
Thus, the vectorial system with incoherent interactions is free of the
instability of soliton chains in single-component models with coherent
(phase-dependent) interactions \cite{instability}.

\emph{Experimental realization.-} The creation of TL supersolitons in BECs
would depend on the use Feshbach resonance 
techniques to get an atomic mixture with attractive
intra-species and repulsive inter-species interactions. Atomic mixtures with controllable interspecies interactions have already 
been reported in Ref. \cite{neww}. Our initial state of alternating solitons may be created by the
modulational instability and segregation from an initially stable
two-component mixture \cite{chinos}.

\emph{Conclusions.-} We have explored a physical model based on the
vectorial NLSE, in which hard-particle-like (bouncing) elastic
collisions between solitons belonging to different species are
possible. These
interactions allow building an analogue of the Newton's cradle using
solitary waves, and to supersolitons in a chain of alternating solitons. The existence of 
these robust localized collective excitations on top of arrays of nonintegrable
solitons represents a remarkable emergent  phenomenon. 

\acknowledgments

\emph{Acknowledgements.-} This work has been partially supported by grants
FIS2006-04190, FIS2004-02466, FIS2007-29090-E (Ministerio de Educaci\'{o}n y
Ciencia, Spain), PGIDIT04TIC383001PR (Xunta de Galicia) and PCI-08-0093
(Junta de Comunidades de Castilla-La Mancha, Spain). D.N. acknowledges a
grant from Conseller\'{\i}a de Educaci\'{o}n (Xunta de Galicia).

\end{document}